\documentclass[12pt]{iopart}
\usepackage{iopams}
\usepackage[latin1]{inputenc}
\usepackage[]{graphicx}
\usepackage{color}
\usepackage[FIGTOPCAP,nooneline]{subfigure}

\begin{document}
\title[]{Reduction of the ordered magnetic moment in YMnO$_3$ with
hydrostatic pressure}
\author{M. Janoschek$^{1,2}$   B. Roessli$^1$ L. Keller$^1$  S.N. Gvasaliya$^{1,3}$}
\address{$^1$Laboratory for Neutron Scattering ETHZ \& Paul Scherrer
Institut, CH-5232 Villigen PSI, Switzerland}
\address{$^2$Physik Department E21, TU M\"unchen, 85748 Garching, Germany}
\address{$^3$on leave from Ioffe Physical Technical Institute, 26 Politekhnicheskaya,
194021, St. Petersburg, Russia}
\author{K. Conder$^4$ E. Pomjakushina$^{4,1}$}
\address{$^4$Laboratory for Developments and Methods, Paul Scherrer
Institut, 5232 Villigen PSI, Switzerland}
\begin{abstract}
YMnO$_3$ exhibits a ferroelectric transition at high temperature
($\approx$~900~K) and magnetic ordering at T$_N \approx$~70~K where
the dielectric constant shows an anomaly indicative of the
magneto-dielectric effect. Here we report powder neutron diffraction
experiments in this compound that show that the magnetic moment at
saturation is reduced by application of hydrostatic pressure. Our
results yield further insight about the nature of the spin-lattice
interaction in ferroic materials.
\end{abstract}
%
%
%
%
\submitto{\JPCM}
%
\maketitle
\section{Introduction}
\begin{figure}[b]
\begin{minipage}[c]{\textwidth}
\centering
\includegraphics[width=10cm]{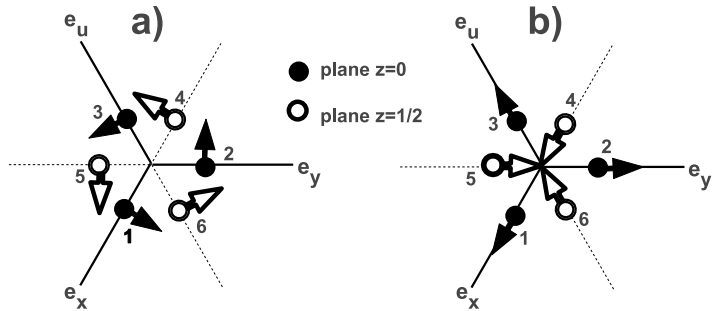}
\end{minipage}
\caption{\small{The two possible magnetic structures for YMnO$_3$
which are in agreement with neutron powder diffraction measurements.
The configuration in the $ab$-plane is shown. (a) Structure defined
by the irreducible representation $\Gamma_1$; the coupling between
the layers $z=0$ and $z=1/2$ is antiferromagnetic. (b) Structure
defined by irreducible representation $\Gamma_3$; the coupling
between the layers $z=0$ and $z=1/2$ is ferromagnetic and for this
structure components of the moments perpendicular to the $ab$-plane
($m_z$) are possible. $m_{iz}=+m_z; i=1,2,3$ and $m_{iz}=-m_z;
i=4,5,6$.}}\label{magstruc}
\end{figure}
Yttrium manganite belongs to the the family of rare-earth manganites
$R$MnO$_3$ ($R =$ rare earth element) which have both ferroelectric
and magnetic order. These compounds crystallize in the hexagonal
space-group P6$_3$cm below the paraelectric-ferroelectric phase
transition (T~$_C\sim$~900~K). The six Mn$^{3+}$ magnetic moments in
the unit cell are located in planes separated by $\sim
5.7$~\AA~along the hexagonal axis. In the $z=0$ plane the Mn atoms
are placed along the $a$ , $b$ and $-(a+b)$ axes, whereas in the
$z=1/2$ plane they are along the axes $-a$ , $-b$ and $(a+b)$. The
Mn atoms are centered in triangular bi-pyramids whose vertices are
oxygen atoms. Because of the large distance between adjacent
triangular layers, YMnO$_3$ is a good candidate for a geometrically
frustrated 2Dimensional antiferromagnet although small exchange
interactions along the hexagonal axis drives magnetic ordering at
$T_N\approx70$~K. The magnetic structure of YMnO$_3$ was first
studied by neutron diffraction by Bertaut \cite{bertaut} who found
that two spin arrangements give similar agreement between observed
and calculated powder diffraction patterns. Both models describe a
triangular arrangement of the $S=2$ magnetic moments in the basal
plane with the coupling between adjacent layers being either
ferromagnetic or antiferromagnetic (further on labeled according to
their irreducible representations $\Gamma_3$ and $\Gamma_1$
respectively; all representations are derived in Ref.~\cite{munoz};
see Fig.\ref{magstruc}). Although both structures lead to very
similar magnetic neutron intensities, Mu\~{n}oz concluded from
recent neutron diffraction studies that the probable magnetic
structure of YMnO$_3$ corresponds to $\Gamma_1$ with the magnetic
moment of the Mn-ions being $\mu=2.9$~$\mu_B$ at
saturation~\cite{munoz}. The reduction of the value of the magnetic
moment from the expected 4~$\mu_B$ of Mn$^{3+}$ free spins is taken
as evidence that even in the ordered phase strong
\begin{table}[t]
\begin{center}
\begin{tabular}{lcll}\hline
&&ambient pressure&9.6~kbar\\\hline $a[\AA]$&&6.15672(5)&6.14676(6)\\
$c[\AA]$&&11.4179(1)&11.4139(1)\\
$v[\AA^3]$&&374.802(4)&373.460(5)\\
atomic positions&&&\\
$Y1$&$z$&0.26989(5)&0.27204(9)\\
$Y2$&$z$&0.22874(3)&0.2309(2)\\
$Mn$&$x$&0.3323(8)&0.337(2)\\
$O1$&$x$&0.31101(1)&0.3170(5)\\
&$z$&0.16069(3)&0.1474(1)\\
$O2$&$x$&0.63937(6)&0.6472(5)\\
&$z$&0.33703(3)&0.3399(1)\\
$O3$&$z$&0.47426(7)&0.4725(3)\\
$O4$&$z$&0.01401(5)&0.0132(3)\\
Agreement factors&&&\\
$\chi^2$&&8.13&5.01\\
Bragg-R-Factor&&1.734&5.085\\\hline
\end{tabular}
\caption{\small{Results of the Rietveld refinement of neutron powder
diffraction data for YMnO$_3$ at 300K for ambient pressure and
9.6~kbar. Atomic positions: Y1 and O3 at 2a $(0,0,z)$; Y2 and O4 at
4b $(\frac{1}{3},\frac{2}{3},z)$; Mn, O1 and O2 at 6c $(x,0,z)$ for
Mn $z=0$}}\label{structural}
\end{center}
\end{table}
spin fluctuations are present due to geometrical frustration \cite{park}.\\
The $R$MnO$_3$ ferroics have received renewed interest since
anomalies in the dielectric constant $\varepsilon$ are found at the
magnetic ordering temperature T$_N$ \cite{huang} indicating a strong
coupling between ferroelectric and magnetic properties. The origin
of the magnetoelectric (ME) effect in these compounds is still not
fully understood but spin-lattice interaction might play an
important role in these materials. For example, anomalies of the
structural parameters were reported at T$_N$ that yield evidence
that coupling between spin and lattice degrees of freedom is at the
origin of the ME effect in YMnO$_3$~\cite{lee,sharma}. In order to
get more insight into the coupling between spin and structural
parameters and their possible relationship with the magnetic and
dielectric properties of ferroic materials, we investigated the
influence of hydrostatic pressure on the magnetic ordering in
YMnO$_3$.

\begin{figure}[th]
\centering\subfigure[]{\label{20K1bar}
\includegraphics[width=12cm]{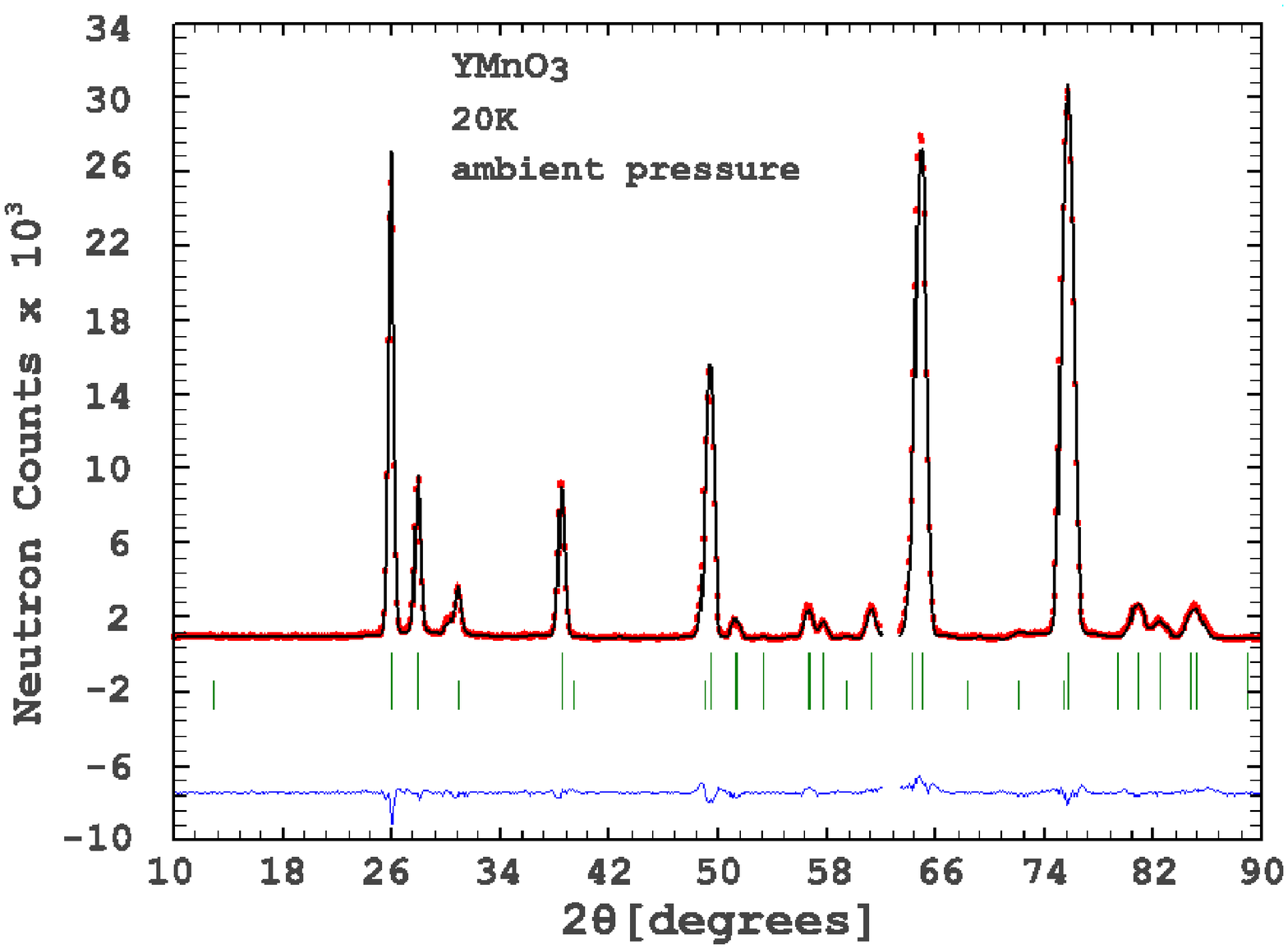}}
\hspace{0.2cm} \centering \subfigure[]{\label{20K7kbar}
\includegraphics[width=12cm]{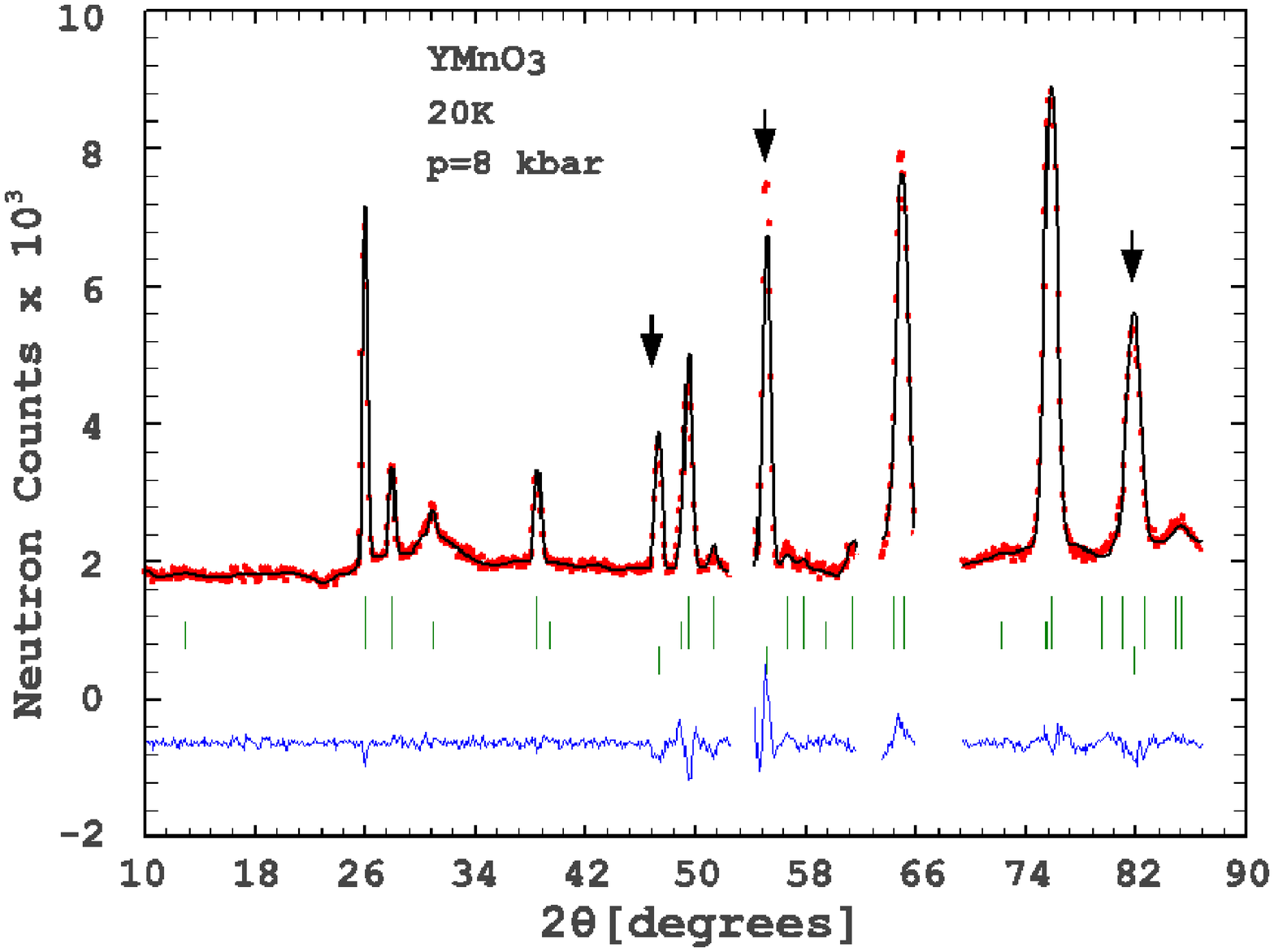}}
\caption{Observed $(\bullet)$ and calculated $(-)$ neutron
diffraction patterns of YMnO$_3$ at T~=~20~K for (a) ambient
pressure and (b) p~=~8~kbar. The bars below the patterns denote the
position of nuclear (upper row) and magnetic Bragg reflections
(lower row). The diffraction pattern for p~=~8~kbar contains
additional reflections due to the NaCl powder (denoted by arrows).
Bragg peaks originating from the pressure cell were excluded from
the data. In addition, we note that the background is modulated
around $2\theta = 28^{\circ}$, mainly due to the presence of
Fluorinert in the neutron beam.}
\end{figure}
\section{Experimental Details}
Polycrystalline YMnO$_3$ was prepared by a solid state reaction.
Starting oxides of Y$_2$O$_3$ and MnO$_2$ with 99.99$\%$ purity were
mixed and grounded followed by sintering at 1000-1200$^0$ C in air
during 100h with several intermediate grindings. Phase purity of the
compound was checked with conventional x-ray diffractometer (SIEMENS
D500). The neutron measurements were performed at the neutron powder
diffractometer DMC located at the cold source of the neutron
spallation source SINQ. The instrument was operated with $\lambda =
2.566$~\AA~ and a PG-filter was installed in the beam to remove
higher-order harmonics. For measurements at ambient pressure, 13~g
of polycrystalline YMnO$_3$
 were filled in a standard vanadium container (${\O}$=15~mm).
For the pressure experiments a clamp-type pressure cell was used,
that can attain a maximum pressure of $15$~kbar. Hydrostatic
pressure is obtained by mixing the powder sample with Fluorinert.
The effective pressure was calculated from the known pressure
dependence of the lattice parameters of NaCl~\cite{menoni} that was
mixed with the YMnO$_3$ sample. The sample was cooled in a $^4$He
cryostat of ILL-type.
\begin{figure}[t]
\begin{minipage}[c]{\textwidth}
\centering
\rotatebox{-90}{\includegraphics[width=8.5cm]{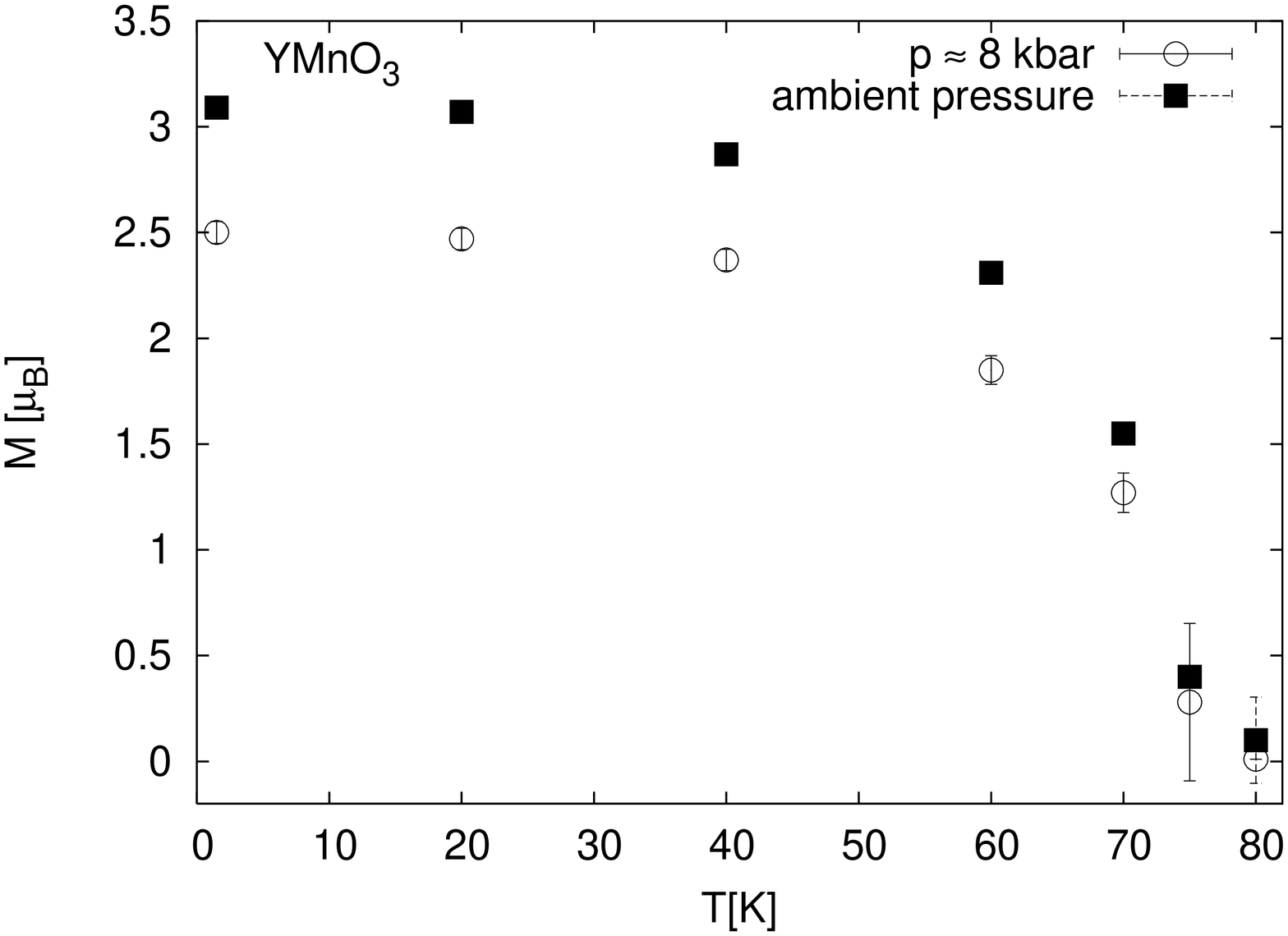}}
\end{minipage}
\caption{\small{Temperature dependence of the magnitude of the $Mn$
magnetic moments for ambient pressure and
p~$\sim$~8~kbar.}}\label{moments}
\end{figure}
\section{Results and Discussion}
A first series of measurements were done at ambient pressure in the
temperature range 1.5~K $<$ T $<$ 300~K. A typical neutron
diffraction pattern is shown in Fig.~2(a) that was analyzed with the
Rietveld method \cite{rietveld} implemented in the program FULLPROF
\cite{rodriguez}. The structural parameters found at T~=~300~K are
presented in Table~\ref{structural}. The fit of the magnetic
structure of YMnO$_3$ was found to yield a slightly better agreement
with the $\Gamma_1$-type (magnetic R$_{Bragg}=4.68$, $\chi^2=5.57$)
than the $\Gamma_3$ structure (magnetic R$_{Bragg}=5.13$,
$\chi^2=5.55$). We note that the symmetry of the $\Gamma_3$
structure allows the magnetic moments to have a component out of the
$ab$-plane. However, agreement factors between observed and
calculated intensities do not improve when spins are canted and we
conclude that the magnetic moments lie in the hexagonal plane. With
the $\Gamma_1$ structure, the value of the magnetic moment at
T~=~1.5~K is $3.09(2)\mu_B$, in good agreement with the results of
Mu\~{n}oz {\it et al.} The temperature dependence of the staggered
magnetization is shown in Fig.~\ref{moments}.

Although the maximum load was applied to the pressure cell, we found
that the effective pressure was 7.8~kbar at T~=~1.5~K and increased
to 9.6~kbar at T~=~300~K. However the pressure increased only
slightly from 7.8 to 8.3~kbar in the temperature range
1.5~K~$<$~T~$<$~80~K where YMnO$_3$ is antiferromagnetically
ordered. Fig.~2(b) shows a typical neutron diffraction pattern
measured at T~=~20~K and p~=~8~kbar. Apart from Bragg reflections of
YMnO$_3$, the diffraction pattern contains now additional
reflections due to the NaCl powder. In addition, the background is
high mainly due to the presence of Fluorinert in the neutron beam.
Therefore the statistical quality of the data is much reduced when
compared with the results obtained at ambient pressure.\\
The applied hydrostatic pressure mainly altered the lattice constant
$a$ whereas $c$ remained almost unchanged ($\Delta a/\Delta
c\approx2.5$). The volume of the unit cell changed from v =
374.802(4)~\AA$^3$~ at ambient pressure to v~=~373.460(5)~\AA$^3$~
at p~=~9.6~kbar. Comparison of the structural parameters shown in
Table~\ref{structural} indicates that only the $z$-coordinate of the
O1 atom is significantly modified by application of pressure. The
shift of the O1 atom decreases the length of Mn-O1 bond from
1.84~\AA~to 1.68~\AA.  A preliminary analysis of the data at T =
1.5~K showed that the neutron diffraction peaks of YMnO$_3$ have
very similar relative intensities than observed at ambient pressure.
Least-square refinements of the diffraction pattern yielded good
agreement factors for the magnetic structure with $\Gamma_1$ and
$\Gamma_3$ symmetry, namely (magnetic) R$_{Bragg}=9.95$ and
$\chi^2=4.73$ for $\Gamma_1$ and (magnetic) R$_{Bragg}=9.97$ and
$\chi^2=4.78$ for $\Gamma_3$ respectively. Hence, the spin
arrangement of YMnO$_3$ is not modified by hydrostatic pressure up
to p~$\sim$~8~kbar. However, the magnitude of the ordered magnetic
moments is significantly reduced from $\mu\sim3$~$\mu_B$ at ambient
pressure to $\mu=2.50(5)~\mu_B$ at p~$\sim$~8~kbar for the
$\Gamma_1$ symmetry. We note, that the ordered moment is reduced to
$\mu=2.44(5)~\mu_B$ for magnetic structure with $\Gamma_3$ symmetry
. The temperature dependence of the ordered magnetic moment is shown
in Fig.~\ref{moments}. Although from neutron powder measurements it
is difficult to determine the value of T$_N$ precisely, we conclude
from our data that the difference of the temperature of the phase
transition from the paramagnetic to the ordered antiferromagnetic
state is less than $\sim$~5~K at p~=~8~kbar as compared to ambient
pressure. A possible model to describe the magnetic and
ferroelectric properties of ferroic materials was proposed by Gong
{\it et al.}~\cite{gong} that includes antiferromagnetic Heisenberg
exchange interactions and a double well potential for the lattice
displacements giving rise to ferroelectricity. The ME coupling is
described by an interaction of the form $g_{\|} u^2_k {\bf
S}_{ai}\cdot{\bf S}_{aj}$ and $g_{\|} u^2_k {\bf S}_{bi}\cdot{\bf
S}_{bj}$ for the intraplane and $g_{\perp} u^2_k {\bf
S}_{ai}\cdot{\bf S}_{bj}$ for the interplane component respectively,
where $g_{\|}$ and $g_{\perp}$ are the intra- and inter-plane ME
coupling constants, $u_k$ is the lattice displacement at site $k$,
and ${\bf S}_{ai}$ is the Heisenberg spin operator, $i$ and $j$
denote the nearest neighbors to $k$ and $a$ and $b$ represent the
nearest planes. In the mean-field approximation, the ME coupling
leads to a renormalisation of the intra- and interplane exchange
integrals $J_{\|}$ and $J_{\perp}$ to $J_{\|}+g_{\|}p^2$ and
$J_{\perp}+g_{\perp}p^2$ respectively, where $p=\langle u_k\rangle$.
In that approach, the N\'eel temperature is reduced by the
ME-coupling, but the value of the magnetic moment $<S^z>$ in the
ground-state is essentially unaffected~\cite{zhong}. Although, this
is consistent with the observation that in HoMnO$_3$, the spin
reorientation temperature T$_{SR}$ of the manganese sublattice
shifts to lower temperatures by about 1.5~K at
p$\sim$8~kbar~\cite{cruz}, the model does not explain the moment
reduction observed in YMnO$_3$.
\section{Conclusion}
We investigated the influence of hydrostatic pressure on the
properties of the magnetic ground-state of YMnO$_3$. The magnetic
structure of YMnO$_3$ is found to have $\Gamma_1$ symmetry at
p~=~8~kbar. On the other hand, the ordered magnetic moment at
saturation is significantly reduced by application of pressure. This
suggests that coupling between the size of the magnetic moment and
the volume of the unit cell is important in ferroic materials and
that application of hydrostatic pressure leads to an increase of the
spin fluctutations in YMnO$_3$ and enhances the 2Dimensional
character of the magnetic properties of that compound. We suggest
that application of pressure leads to a renormalisation of the
temperature dependence of the staggered magnetisation, i.e. the
strain influences the sublattice magnetisation~\cite{mattis}.

\section{Acknowledgments}
We thank D. Sheptyakov for help and advice with the pressure cell
and P. J. Brown for fruitful discussions. This work was performed at
the spallation neutron source SINQ and was partially supported by
NCCR MaNEP.
\newline

\end{document}